# Hot-carrier optoelectronic devices based on semiconductor nanowires

**Jonatan Fast[1,a)], Urs Aeberhard[2], Stephen P. Bremner[3], Heiner Linke[1,a)]**

[1]NanoLund and Solid State Physics, Lund University, Box 118, 22 100 Lund, Sweden

[2]Integrated System Laboratory, ETH Zurich, 8092 Zürich, Switzerland

[3]School of Photovoltaic and Renewable Energy Engineering, UNSW Sydney, Sydney 2052, Australia

a)Author to whom correspondence should be addressed: jonatan.fast@ftf.lth.se, heiner.linke@ftf.lth.se

**ABSTRACT**

In optoelectronic devices such as solar cells and photodetectors, a portion of electron-hole pairs are generated as so-called hot carriers with an excess kinetic energy that is typically lost as heat. The long standing aim to harvest this excess energy to enhance device performance has proven to be very challenging, largely due to the extremely short-lived nature of hot carriers. Efforts thus focus on increasing the hot carrier relaxation time, and on tailoring heterostructures that allow for hot-carrier extraction on short time- and length-scales. Recently, semiconductor nanowires have emerged as a promising system to achieve these aims, because they offer unique opportunities for heterostructure engineering as well as for potentially modified phononic properties that can lead to increased relaxation times. In this review we assess the current state of theory and experiments relating to hot-carrier dynamics in nanowires, with a focus on hot-carrier photovoltaics. To provide a foundation, we begin with a brief overview of the fundamental processes involved in hot-carrier relaxation, and how these can be tailored and characterized in nanowires. We then analyze the advantages offered by nanowires as a system for hot-carrier devices and review the status of proof-of-principle experiments related to hot-carrier photovoltaics. To help interpret existing experiments on photocurrent extraction in nanowires we provide modeling based on non-equilibrium Green's functions. Finally, we identify open research questions that need to be answered in order to fully evaluate the potential nanowires offer towards achieving more efficient, hot-carrier based, optoelectronic devices.

## Contents

## I. INTRODUCTION

When a semiconductor material is illuminated with light of energy higher than its bandgap, the excess energy is given to the photoexcited electron and hole in the form of kinetic energy (Fig. 1 (a)). Such carriers with excess energy are referred to as hot carriers. Typically, the hot carriers relax towards the band edge within picoseconds, transferring their kinetic energy into heat via various scattering events involving lattice vibrations (phonons) and interaction with other carriers. The processes immediately following absorption, on timescales down to femtoseconds, is a topic of many studies, both experimentally and theoretically.[1–3]

The excess kinetic energy of hot carriers is recognized to offer opportunities for a variety of applications. In chemistry, this includes new reaction pathways, improved photochemical reactions, and artificial photosynthesis.[4,5] Excitation of hot carriers in metallic nanostructures provides a means of local heating with high precision,[5] applied for example in pioneering cancer treatment methods[6] and growth of nanostructures.[7,8] In metal-oxide-semiconductor field effect transistors on the other hand, the generation of hot carriers by strong electric fields has been known for over 50 years to have degrading effects.[9] Photodetectors operating in the infrared regime are being developed based on the principle of exciting hot carriers in metal (plasmonic) structures, and injecting them into semiconductor structures.[10–14] There is also significant interest in harvesting hot carriers to make use of their excess energy in so called hot-carrier photovoltaic (HCPV) devices. Specifically, it is a major aim to harvest the energy from hot carriers in order to boost the efficiency of current solar cell technology beyond the conventional limits. However, no HCPV device has of yet been realized that even comes close to competing with the efficiency of standard solar cell technology.[15]

Lately, intriguing reports have emerged suggesting that the dynamics of hot carriers might be altered in semiconducting nanowires, due to their shape and dimensionality.[16–18] A nanowire is defined as a rod like structure with a high length/diameter aspect ratio, and a diameter on the order of 1-100 nm. Nanowires have been the subject of intense research interest, due to their novel and tailorable optical, electronic, and thermal properties,[19] and because nanowire synthesis provides a platform for flexible and high-quality band-engineering.[20–22] Currently, nanowires are being considered for a wide range of applications such as transistors,[23] light emitting diodes,[24] solar cells,[20] quantum computing,[25,26] and thermoelectrics,[27,28] to mention a few.

This paper aims to review the current state of experiment and theory of the hot-carrier regime in semiconductor nanowires, and how nanowires might be exploited in the future to create new classes of devices. For the less familiar reader, we will start by giving a brief introduction to the field of hot-carrier photovoltaics in Section II. In Section III we introduce the main processes involved in the relaxation dynamics of hot carriers, and discuss how these processes might be different in nanowires compared to bulk. Section IV reviews the current literature on experimental measurements of the hot-carrier dynamics in nanowires. Practical approaches towards realizing nanowire-based HCPV devices are then explored in Section V. Specifically, we will focus on hot-carrier excitations taking place directly within semiconductor nanowires, in contrast to injection from metal structures.[10] Finally, we summarize the main conclusions with regards to the prospect of utilizing nanowires in future HCPV devices. We also identify what we consider to be key open questions to the field of HCPVs, offering guidance for future efforts.

## II. HOT-CARRIER PHOTOVOLTAICS

In this section we give a brief overview of the concepts, history, and current state of the field of HCPVs. This will serve to motivate and put this article into context. We will introduce the two main concepts of consideration when designing a HCPV device, namely 1) the maintenance and 2) the extraction of a hot-carrier population, both of which will be discussed in detail in the later sections. In no way is this attempting to be an exhaustive review of HCPVs, and for further reading on that topic the reader is referred to excellent, existing review articles,[1,3,29–32] most recently by König et al.[15].

In the photovoltaic (PV) power conversion process, there are a number of fundamental losses that limit the power conversion efficiency for a single band gap device to approximately 33% under 1 sun illumination (Fig. 1(b)), and to 40% under maximum light concentration. These limits are well known within the PV field as the Shockley-Queisser limits, stemming from the seminal detailed-balance approach presented in 1961.[33] Five different types of fundamental loss mechanisms have been identified by Hirst and Ekins-Daukes[34] as:

1. *Carnot loss* stems from the fundamental Carnot efficiency limit to any energy or power conversion process that behaves as a heat engine, corresponding to a completely reversible process. In the case of photovoltaic power conversion, this limit is set by the temperature of the sun (the heat source), $T_{sun}$ = 5778 K, and that of the converter, typically assumed to be at room temperature $T_{room}$ = 300 K. The Carnot limit is then $(T_{sun} - T_{room})/ T_{sun} \approx 95\%$.

2. *Sub-band gap loss* exists because only photons with energy larger than the energy band gap contribute to the generation of electron-hole pairs (Fig. 1(a)). The energy of photons within the solar spectrum that have energy below the band gap are thus lost.

3. *Boltzmann loss*, also referred to as étendue loss, is the entropic loss associated with the difference between the possible states for an incident cone of light and the cone of emitted light from the surface.

4. *Emission loss* represents the carriers lost (recombined) before collection and extraction. In the so called radiative limit (only radiative recombination present) this equates to the emission photon flux from the solar cell, set by the chemical potential difference between electrons and holes in the solar cell.

5. *Carrier relaxation loss*, refers to the excess energy a hot electron (hole) has above (below) the conduction (valence) band edge (Fig. 1(a)). Because this excess energy is lost to heat in traditional solar cells, this loss is referred to as thermalization loss in the photovoltaics literature. In this review, to stay consistent with the hot-carrier literature, we will refer to this loss as *carrier relaxation* loss. The term *thermalization* will be reserved for inelastic carrier-carrier scattering events, as discussed in Section III.

In addition to these fundamental effects, one needs to consider device-level effects such as parasitic resistances, transport losses or imperfect absorption.

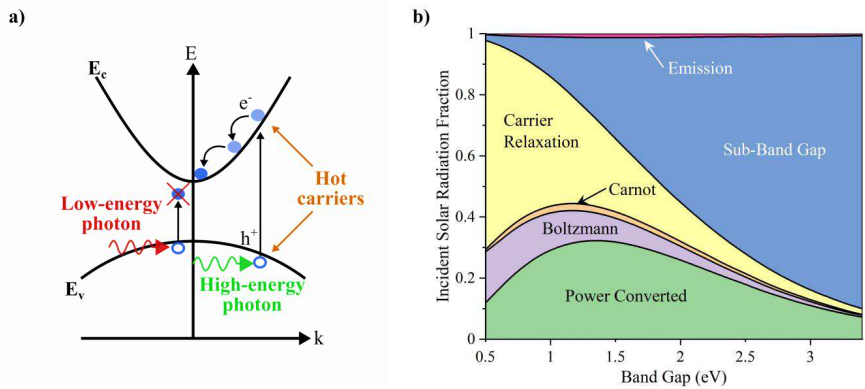

Fig. 1. (a) Simplified band-diagram picture of optical excitation in a semiconductor, indicating two sources for loss in efficiency: (i) the inability to absorb sub-bandgap photons, and (ii) the relaxation of hot carriers to the conduction/valence band edge. In the example shown here the effective mass of the holes is larger than that of the electrons. Momentum conservation then requires that a majority of the excess photon energy is given to the electrons. (b) The five fundamental losses and their relative impact on the conversion efficiency in a standard single junction solar cell, as a function of the absorber material bandgap. *Adapted from ref.[35] with permission from the authors.*

Comparing the relative impact on conversion efficiency for each of the five fundamental losses, the two largest limiting losses are due to sub-bandgap photons and hot-carrier relaxation (Fig. 1(b)).[34] Reduc-ing the band gap to improve efficiency has limits as there is a trade-off with increasing carrier relaxation losses. This trade-off can be partially circumvented by using multi-junction approaches, where a series of band gaps is used in order to reduce the carrier relaxation losses for portions of the spectrum. There are technical limitations as to how far this approach can be taken, but currently the champion efficiency for such PV conversion sits at 47.1%.[36,37] In HCPVs, the idea is instead to use a low bandgap absorber material, and create a device that extracts the hot carriers at energies above the bandgap (a more detailed description is presented in Section V). The idea was first proposed by Ross and Nozik who in 1982 theoretically predicted that the efficiency of such a device could reach 66%.[38] By taking into account carrier-carrier scattering processes, Würfel expanded on their model and predicted a maximal efficiency of 85% under concentrated light.[39] This highlights the fact that an important aspect of HCPVs is the ability of hot carriers to scatter inelastically and redistribute their energy amongst each other to form a thermal distribution, i.e. thermalize.[40]

Alternatively, in a process referred to as multiple exciton generation (MEG), hot carriers with an excess energy larger than the absorber bandgap excite an additional electron-hole pair via impact ionization when they relax. In the MEG approach, the goal is to reach external quantum efficiencies above 100% and thus a photocurrent above the Shockley-Queisser limit, in contrast to HCPVs where the goal is a higher photovoltage. In this review our focus will be on the latter, and for work on MEG in various nanostructures we refer to existing reviews.[41,42]

A key challenge in the realization of a HCPV device is the inherent nature of hot carriers to relax very rapidly (order of ps)[3]. Efforts therefore typically focus on one of two aspects: slowing down the relaxation process of hot carriers; and designing contacts that can extract hot carriers at elevated temperatures/energies, before they lose their heat to the lattice. To slow down the relaxation, various ways of altering the scattering rates involving carriers and phonons are studied. To alter the nature of these scattering events, candidate systems typically include intricate band-engineering, quantum confinement, and high surface/volume ratios or super-lattice structures. Already in the 80's and 90's, a large body of hot-carrier luminescence studies was performed on various types of quantum well structures, mainly based on GaAs/AlGaAs.[43–49] Increased hot-carrier temperatures and slower cooling rates were found in these systems compared to bulk, but only at high excitation densities (one study found the critical density of hot carriers to be on the order of $10^{18} cm^{-3}$)[49]. In more recent years, there have been a number of studies exploring the carrier-phonon dynamics in various quantum-well systems,[50–54] as well as in bulk materials whose phononic and electronic band structures are predicted to lead to increased carrier relaxation times.[55–60]

Realization of a HCPV device will likely require a precise control of fabrication at the nanometer-scale. First, because significant relaxation typically occurs over distances on the order of some 100 nm (Table 1), implying that carrier extraction must take place very close to the location of photoexcitation. Second, because extraction of hot carriers requires contacts with energy selectivity (to harvest only hot carriers), as well as carrier selectivity (to separate electrons from holes). There are a number of experimental demonstrations of nanostructures that extract hot carriers. Extraction of hot carriers through resonant tunneling contacts have been demonstrated in GaAs/AlGaAs,[61,62] Si quantum dots on $SiO_2$,[63–66] and defects in insulators.[67] A hot-carrier photocurrent was generated in InP/PbSe quantum wells,[68] InGaAs quantum wells,[69] and InAs/GaAs quantum dot super lattices.[70] Ultrafast hot electron extraction has been observed from PbSe nanocrystals to a $TiO_2$ acceptor contact.[71] High photovoltages have also been observed in ferroelectric insulators,[72,73] though it appears to be at the cost of low current and efficiency. In recent years, the prospect of using semiconductor nanowires for HCPV devices has emerged, and in 2017 the demonstration of a photovoltage above the Shockley-Queisser limit was achieved in InAs/InP heterostructure nanowires.[74] In this review, our focus is on semiconductor nanowires and their potential benefits for HCPV devices.

## III. THE HOT-CARRIER RELAXATION PROCESS

### III.A. General

In this section we will give an overview of the relaxation processes involved in the energy dissipation of hot carriers, and how these are expected to be altered in nanowires. Energy dissipation due to hot-carrier relaxation is the result of a complex chain of processes related to the mutual interaction of charge carriers and lattice vibrations that are driven out of equilibrium. The sequence of these processes starts with the generation of a non-equilibrium occupation of highly excited states. Under certain conditions, this population evolves quickly into a thermal distribution at elevated carrier temperature, which cools down due to coupling to lattice vibrations. Finally, the thermal energy is dissipated via propagation of heat, in the form of acoustic phonons, away from the location of generation. These processes are illustrated schematically in Fig. 2(a) and the order of magnitude of the typical time scales and distances are given in Table 1. It should be pointed out that the relative importance, as well as the time scale, of the different relaxation processes may vary greatly depending on parameters such as material and carrier density. As such, the values in Table 1 should be taken simply as guidelines to be used for comparison and discussion. In this section, we aim only to give an introduction to the relaxation processes which will be discussed in later parts of the article. For a more in-depth discussion we refer the reader to other literature, such as the recent review by König *et al.*[15]

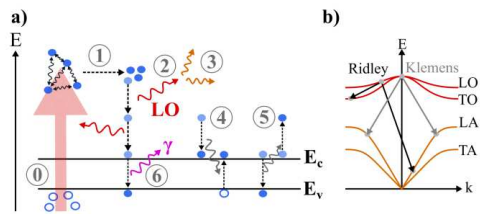

Fig. 2. (a) The process of excitation and relaxation of hot carriers, shown for electrons. 0) Excitation, 1) Carrier-carrier scattering to form a *thermalized* distribution with a distinct carrier temperature $T_C$, 2) Carrier relaxation via interaction with optical phonons, 3) Optical phonon decay into acoustic phonons, 4) Impact ionization, 5) Auger recombination, and 6) Radiative recombination. (b) The symmetric (Klemens) and asymmetric (Ridley) decay channels for LO phonons. For time- and length-scales, see Table 1.

### III.A.1. Thermalization of hot carriers

(Opto-)electronic semiconductor device operation imposes inherently non-equilibrium conditions to the charge carriers in the device. Under external application of bias illumination or voltage, energy is transferred to the carriers, which undergo transitions to excited states. Illumination with light of photon energy exceeding the band gap energy of the semiconductor and acceleration in strong internal fields both lead to the formation of a carrier population that is no longer in equilibrium with the ionic lattice. At turn-on of the excitation, the resulting population reflects primarily the excitation mechanism as well as the initial and final states participating in the transition. Depending on the excitation strength and mechanism, only a fraction of the carriers are excited, and the non-equilibrium charge carrier population coexists with a "cold" charge carrier population that is in thermal equilibrium with the lattice.

In the presence of sufficiently large density, excited charge carriers may redistribute their kinetic energy via carrier-carrier scattering processes amongst each other according to Fermi-Dirac statistics, a process that tends to occur on a time scale of 10-100 fs.[29] In the absence of an electric field, the resulting carrier population corresponds to a local quasi-equilibrium at a defined quasi-Fermi level and temperature, which is why the associated process is termed *thermalization*. The temperature of this thermal distribution will be referred to as the carrier temperature, $T_C$, and it can be substantially higher than that of the lattice, $T_L$. Sustaining a hot-carrier population that may be used for applications will require $T_C > T_L$ in a steady state situation. In the case where e-h pairs are generated, the temperatures of the excited electron and hole populations can be different, depending on the effective mass ratio. The effective mass ratio determines the distribution of the excess energy (Fig. 1(a)), in that the lighter carrier type gets a larger part of the excess energy. The temperatures of the excited electron and hole populations is also affected by the presence of electron-hole scattering processes such as impact ionization (Fig. 2(a)), which couple the two populations, but require excitation energies in excess of twice the band gap energy. It should be stressed again that in the photovoltaics literature the term thermalization often refers to the complete process whereby hot carriers release all their excess kinetic energy to lattice heat until $T_L = T_C$, differing from the use of this term in most hot-carrier literature, and in this article.

| Process | Time scale, $\tau$ | Length scale, $l$ | |
| --- | --- | --- | --- |
| | | $m_{eff} = 1 \cdot m_e$ | $m_{eff} = 0.05 \cdot m_e$ |
| 0. Excitation[15] | Instantaneous | | |
| 1. Thermalization, carrier-carrier[29] | $10^{-13} - 10^{-14}$ s | 1 - 10 nm | $5 \cdot (10 - 10^2)$ nm |
| 2. Carrier-phonon[3] | $10^{-12}$ s | $10^2$ nm | $5 \cdot 10^2$ nm |
| 3. Optical phonon decay[2,3] | $10^{-11}$ s | | |
| 4. Impact ionization[75] | $10^{-11} - 10^{-12}$ s | $10^2 - 10^3$ nm | $5 \cdot (10^2 - 10^3)$ nm |
| 5. Auger recombination[3] | $10^{-10}$ s | $10^4$ nm | $5 \cdot 10^4$ nm |
| 6. Radiative recombination[3] | $\geq 10^{-9}$ s | $\geq 10^5$ nm | $\geq 5 \cdot 10^5$ nm |

Table 1. Typical timescales for the relaxation processes displayed in Fig. 2(a). The typical length scale, $l$, for each process is estimated by $l = \tau \cdot v_{th}$, where $v_{th}$ is the thermal drift velocity, estimated from $(m_{eff} v_{th}^2)/2 = 3kT/2$ (i.e. 3D), where $m_{eff}$ is the carrier effective mass, $k$ the Boltzmann constant and $T$ the ambient temperature set to 300 K. $l$ is shown as estimated from two different effective electron masses of $m_{eff} = 1 \cdot m_e$ and $m_{eff} = 0.05 \cdot m_e$ ($m_e$ being the standard electron mass), as most semiconductors lie roughly in this range. For example, $m_{eff} = 0.82 \cdot m_e$ for electrons InP, and $m_{eff} = 0.023 \cdot m_e$ for electrons InAs.

### III.A.2. Hot-carrier relaxation via lattice vibrations

In a sense what drives the further relaxation of the hot carriers is the equilibration of the two temperatures $T_C$ and $T_L$. This process is in this article referred to as *hot-carrier relaxation*. Relaxation is typically mediated by interaction with lattice vibrations (phonons), or via impact ionization and Auger processes that transfer energy between electron and hole populations.

In polar semiconductors, hot carriers mainly release their energy in the form of longitudinal optical phonons (LO), with a specific energy, via electrostatic Fröhlich type interactions.[75,76] The emission of LO phonons typically occur on the order of 1 ps,[3,75] but this time scale depends on the material-dependent carrier-phonon scattering phase space and the coupling strength between LO phonons and carriers. The hot carrier will subsequently approach the band edge via a series of phonon emissions.

In principle, the carriers can also *reabsorb* phonons, gaining energy, as first observed by Shah *et al*.[77]. The rate of reabsorption depends on the distribution of phonons: in the presence of a finite optical phonon lifetime, the phonon population is not given by the thermal occupation at the lattice temperature, but depends on the net phonon emission rate and the subsequent rate of phonon decay due to anharmonic scattering. The build-up of a non-equilibrium phonon population, and the associated reabsorption of phonons by charge carriers, can be observed as an effective increase of hot-carrier life time. Such observation is usually described as a *hot phonon bottleneck* effect,[77] even if the phonon distribution is not thermal.

The strength of electron-phonon interaction depends on the carrier density in a non-trivial way. On the one hand, electron-phonon interaction requires the presence of carriers with excess energy larger than the optical phonon energy. On the other hand, the electron-phonon coupling strength is reduced at higher carrier density due to *screening*. Elevated carrier densities may also be accompanied by the emergence of plasmons (coherent carrier oscillations) to which the optical phonons may couple, and the energy loss rate is then determined by the plasmon damping.[78]

### III.A.3. Phonon decay and heat dissipation

The phonon modes considered in the electron energy loss process are the solutions of the *harmonic approximation* of the phonon Hamiltonian, which considers the expansion of the ionic lattice potential to second order in deviations from the equilibrium lattice positions. Higher order components in this Taylor expansion are termed as *anharmonic*, and are at the origin of phonon scattering processes, inducing the finite life time of the phonon modes.[79,80] For instance, the third order component gives rise to three-phonon processes, in which an initial high energy mode decays into two lower energy modes under observation of energy conservation.

There are many different forms of anharmonic phonon decay in semiconductors, depending on the phonon dispersion, i.e., the energy vs momentum relation of longitudinal and transverse optical (LO, TO) and acoustic (LA, TA) phonon modes (Fig. 2(b)). In many bulk semiconductors, the LO phonons generated by the hot carriers decay into lower energy phonons via either the symmetric Klemens[81] decay (LO → LA + LA) or the asymmetric Ridley[82] decay (LO → TO + LA). Both processes preserve energy and momentum. Fig. 2(b) presents a simplified quasi-1D version of these processes: in reality, the anharmonic interaction can couple many different modes, as long as the energy conservation and the selection rules imposed by the matrix element of the scattering process are fulfilled.[15]

The Klemens decay channel is inhibited if the phonon dispersion contains a band gap larger than the maximum energy of the acoustic phonons. This can result in a slowing down in the decay of LO phonons, and for example provides an explanation for the observation of longer LO lifetimes in InP compared to GaAs, both in bulk[83] and nanowire[84] shape. Irrespective of the decay mechanism, LO phonons eventually decay into acoustic phonons, typically on timescales of the order of 10 ps.[2,3]

In general, the acoustic modes themselves couple only weakly to carriers,[85] although their contribution should be considered in a quantitative assessment of the energy loss rate.[86] On the other hand, they are responsible for the transport of heat away from the source. If acoustic phonons are generated (i.e. via optical phonon decay) faster than they equilibrate with the "cold" acoustic phonons at $T_L$, then a build-up of non-equilibrium acoustic phonon population may arise, sometimes referred to as an *acoustic phonon bottleneck*.[87,88] Such an effect was first experimentally observed in 1995 by Klimov *et al*.[89], and is expected to inhibit the decay of optical to acoustic phonons.[87,88] A build-up of optical phonons may in turn lead to an increased carrier lifetime, and thus an increased $T_C$. Only in special cases acoustic phonons might be up-converted to optical ones, which contributes to the hot phonon bottleneck effect.[90] Therefore, the hot-carrier energy is typically considered to be lost once the optical phonons have been converted to acoustic modes.

### III.B. Nanowires

Nanowires have only recently been shifting into the attention of hot-carrier device research, and little nanowire-specific data on hot-carrier physics is available. However, some traits of nanowires can be discussed in the light of their impact on hot-carrier relaxation by consideration of the results obtained in other nanostructure systems.

For most semiconductor materials, electronic confinement is relevant only in very thin nanowires (diameters below 30 nm), such as the structures used for nanowire transistors. There, transverse subband quantization affects the energy dependence of the magnitude of electron-phonon interaction via the density of electronic states, which decreases towards larger energies.[91]

Lateral confinement in wider nanowires, where electronic (quantum) effects are not expected, may nevertheless have advantages over bulk systems, because diffusion in nanowires is inherently directed along the axial direction. In a fully 1D system (wire diameter comparable to the electron wave length), this is known to be the case,[92] however at which diameters this ceases to be the case is, to our knowledge, not clear at the moment. Potentially, the rectification of carrier momentum which is required for a bulk absorber (where carriers are generated with a random momentum distribution, as discussed by König et al.[15]), may not be needed in a nanowire.

Phonon confinement effects emerge at relevant structure sizes comparable to the phonon coherence length, which is on the order of 10 nm for acoustic modes.[15,93,94] Hence, phonon mode quantization is again most relevant in ultra-scaled devices, and was shown to have a sizable impact on carrier relaxation rates in thin III-V quantum wires (Section IV.D).[93,95,96] This is similar to the situation in quantum wells, where confined phonon modes are believed to promote the hot phonon bottleneck effect.[2] If the material allows for acoustic phonon up-conversion, such effect will benefit from the reduced acoustic phonon propagation observed in nanowires.[97]

Most existing work on phonons in nanowires is concerned with the thermal conductivity due to the propagation of acoustic phonons in the presence of various scattering mechanisms, such as anharmon-ic/Umklapp, mass-difference, boundary and electron-phonon scattering (Section IV.C).[98] Explicit consideration of confined phonon modes reveals substantially suppressed heat conductivity as compared to bulk. This is a result of reduced phonon density of states and group velocity,[97] as well as modified anharmonic decay,[99] in addition to boundary scattering.

For length scales above the regions where phonon confinement is relevant - where most of the systems designed for optoelectronic applications are situated - the effect of the wire geometry on phonons is mostly through boundary scattering.[100] Indeed, deviations from bulk behavior in terms of thermal conductivity can be observed up to relative large nanowire diameters (Section IV.C). Interestingly, there are crossovers between materials explained by a change in the dominant phonon scattering mechanism from anharmonic to boundary when going from bulk to small diameters.[101]

The quasi-1D nature of nanowires have also been theorized to lead to increased Coulombic interactions among excitons,[102] in agreement with observations of an increase of MEG in PbSe nanorods compared to both bulk and quantum dots.[103,104] Nanorods are essentially nanowires with a lower length-to-diameter aspect ratio, and proof-of-principle solar cells with PbSe nanorods have been demonstrated to achieve external quantum efficiencies above 100%.[105] There appears however to be an optimum length-to-diameter aspect ratio for PbSe nanorods of about 6-7, above which the rate of MEG quickly decreases, suggesting that nanowires, with their higher aspect ratio, may not yield high levels of MEG.[103]

The high surface/volume ratio of nanowires makes the carrier dynamics highly susceptible to impact from impurities at the surface, an issue addressed briefly in Section V.D.

Finally, opto-electronic applications of nanowires are largely based on photonic confinement effects at subwavelength diameters (Section V.C). Such spatial confinement of carrier generation can lead to high local carrier densities, which is the main requisite for most hot-carrier effects.

## IV. EXPERIMENTS ON HOT-CARRIER DYNAMICS IN NANOWIRES

In this section, a survey of experimental results probing hot-carrier dynamics and the phonon characteristics in nanowires is presented. Results across a wide range of materials indicate that hot-carrier and phonon dynamics in nanowires may differ markedly from their bulk counterparts. We will start by reviewing ultra-fast time resolved measurements that probe the hot-carrier relaxation process, followed by steady state measurements. A modified phonon spectrum will not only have a direct impact on the hot-carrier relaxation process but will also impact thermal conductivity. Here we discuss thermal conductivity with a focus on modified acoustic phonon dynamics, and how this may affect the hot-carrier relaxation process overall. Finally, experiments probing the electron-phonon interaction will be reviewed. It should be noted that in several of these experiments, tracking of $T_c$ is done under the assumption that no electric field is present, in order to assume that a thermal distribution is formed.

### IV.A. Ultra-fast probing of carrier relaxation

Ultra-fast measurements, probing carrier dynamics immediately following light absorption, offer powerful techniques for determining the mechanisms by which excited carriers proceed to form a thermal distribution and eventually recombine. The analysis of such measurements is, however, complicated by the fact that the timescales of the various processes of interest, such as carrier-carrier and carrier-phonon scattering, partially overlap.[106] Another point to note is that the carrier densities in these laser based experiments are typically more than three orders of magnitude larger than the highest carrier densities expected for solar energy conversion applications. Effects such as the Moss-Burstein shift of the band gap (bleaching), as well as Pauli blocking or carrier relaxation due to a low density of available final states, can make interpretation challenging. A fuller description of these issues can be found in König *et al.*[15] In short, care must be exercised when interpreting the results of these experiments, particularly in relation to a device operating under steady state conditions.

Ultrafast studies of hot-carrier dynamics in nanowires have mainly been performed by transient Rayleigh scattering, where an initial pump pulse excites a hot-carrier distribution and a subsequent probe pulse is used to measure the change in Rayleigh scattering from the nanowire. The carrier density and temperature are thereby indirectly probed at various time delays, and the carrier energy loss ratio can be fitted as a sum of contributions from emission of optical and acoustic phonons (as shown for GaAs and InP in Fig. 3(a) and Fig. *3*(b)).

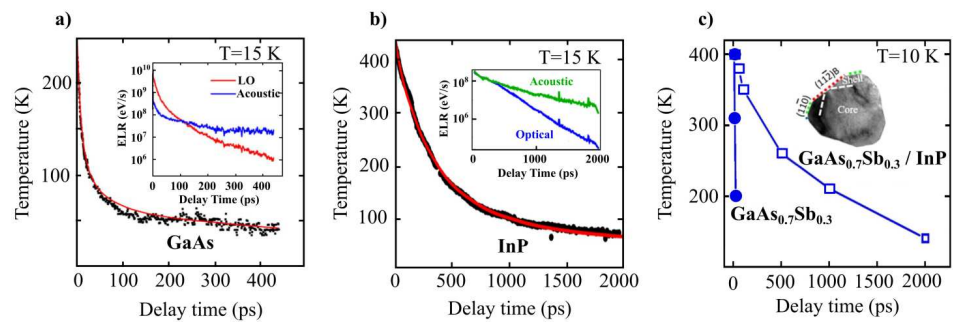

Fig. 3. Electron-hole plasma (EHP) temperature as function of time from excitation, extracted by transient Rayleigh scattering on (a) ZB GaAs nanowire, (b) ZB InP nanowire, and (c) GaAsSb (filled circles) and GaAsSb-InP core-shell nanowire (open squares). The inset of (a) and (b) shows the energy loss rate (ELR) versus time, related to decay of acoustic and optical (LO) phonons respectively. *(a) Reproduced (adapted) with permission from Nano Lett. 12, 5389 (2012). Copyright 2012, American Chemical Society (b) Reproduced (adapted) with permission from Nano Lett. 14, 7153 (2014). Copyright 2014, American Chemical Society (c) Reproduced (adapted) with permission from Nano Lett. 19, 5062 (2019). ref.[18] Copyright © 2019, American Chemical Society.*

Using transient Rayleigh scattering, it was found that in GaAs nanowires the energy loss rate is initially dominated by emission of optical phonons. However, after the initial optical phonon-driven cooling, the rate comes to be dominated by acoustic phonon emission on longer timescales.[107] In InP nanowires, on the other hand, the energy loss rate due to optical phonon emission is initially on the same order as that due to acoustic phonon emission, and optical phonon emission never dominates.[84] Acoustic phonon emission is seen to be the dominant contribution to the energy loss for InP nanowires throughout the entire decay process, resulting in a significantly slower decay of hot carriers in InP compared to GaAs nanowires (a sign of suppressed optical phonon emission). This difference is also seen in bulk samples, where the LO phonon lifetime is reported to be on the order of 3-4 times higher in InP than in GaAs,[108] and can be explained by the phonon band gap in InP[109] inhibiting the Klemens decay of LO→2LA. It should be noted that some differences exist between the two experiments on GaAs and InP nanowires shown in Fig. 3(a) and Fig. 3(b)). Whereas the initial excitation density was on the order of $10^{18}$ cm$^{-3}$ and the temperature was 15 K in both cases, the InP nanowires had a diameter around 150 nm while the GaAs nanowires were roughly 65 nm. Additionally, the GaAs nanowire contained a 10 nm shell of AlGaAs and a capping layer of 5 nm GaAs in order to prevent oxidation, while the InP sample has no mention of surface passivation.

The addition of an InP shell surrounding a GaAsSb nanowire has been shown to greatly increase the life time of hot carriers, as can be seen in Fig. 3(c).[18] The energy loss rate due to optical phonons was found to decrease by an order of magnitude at 300 K for nanowires with the InP shell (and more than two orders at 10 K). While this is indeed a noticeable change, it remains unclear if this increase is due to extra functionality enabled by the shell itself, or due to hot-carrier lifetimes being typically longer in InP than in GaAs, assuming the GaAsSb properties are close to those for GaAs. Possibly, phonons could be confined to the shell, as strong effects of interface phonon modes are already known to exist in quantum wells.[2] Furthermore, the mismatch in dielectric/elastic material constants at the interface will also affect the overall phonon dispersion. To gain further insight about the impact of shell layers on the hot carrier dynamics in nanowires, further studies of such samples would be of great interest.

### IV.B. Steady state probing of carrier relaxation

Under continuous illumination, steady state rates for all the mechanisms involved in hot-carrier generation and relaxation will be formed (see Section VI for a summary of the rates involved). Steady state measurements can thus be seen as a bridge between the ultra-fast techniques (exploring fundamental physics of the hot-carrier relaxation process), and the operation of a device. Steady state experiments allow for example to measure and map the hot-carrier temperature differential, $\Delta T_C = T_C - T_L$, in a sample by studying the continuous wave photoluminescence spectrum of a sample. $T_C$ can be extracted by fitting a modified Planck equation[50,110] or Maxwell –Boltzmann[51,111,112] term to the emitted intensity as a function of energy for the high energy tail (energies above the band gap and Fermi level). This assumes that thermalized carrier distributions have been formed, and that the emissivity (or equivalently the absorptivity) is constant with energy.[51]

Using continuous wave micro-photoluminescence (µ-PL) Tedeschi et al.[16] were able to show a drastic and systematic increase of $\Delta T_C$ with decreasing diameter, $d$, in both InP and GaAs nanowires with various crystal morphology and growth techniques (Fig. 4.). A $\Delta T_C$ of up to 190 K was observed for the case of a 61nm diameter InP nanowire under illumination with a 532 nm laser. Strikingly, with $T_L$ kept constant at 290 K, the trend of increasing $T_C$ with decreasing diameter (i.e. $\Delta T_C \sim d^{-1}$) is seen for nanowires with a diameter from 350 nm downwards. A $\Delta T_C > 100$ K can be seen for nanowires of 100 nm diameter or less, well above the diameters where phonon and electron confinement effects are expected to be significant (Section III.B). Interestingly, it was found that the $\Delta T_C$ values extracted for a particular diameter of nanowire did not change upon increase of the illumination power density, an observation not readily explained at this point of time. It may however indicate that in a steady-state situation it is very challenging to achieve the high carrier and phonon densities required to invoke hot phonon bottleneck effects. With the relatively low power densities used in the study, Auger effects could be neglected. Time resolved µ-PL measurements further established that the µ-PL characteristic decay times are much longer than expected for carrier relaxation times. Since hot phonon effects could be ruled out, and the effects were independent of growth method and crystal structure (WZ or ZB), this leads the authors to the conclusion that phonon boundary scattering is the most likely reason for the diameter dependence seen in these results. The study also found that $\Delta T_C$ increased exponentially as a function of $T_L$. As a higher $T_L$ is accompanied by an increase of acoustic phonon density, this might in the end have an effect of increasing $\Delta T_C$, a possibility we will discuss more in section IV.C.

The achievement of decoupling between $T_C$ and $T_L$ under steady state illumination has also been observed in Si nanoribbons.[82] The authors observed an increase in carrier relaxation time for higher $\Delta T_C$, accredited to a reduction in electron-phonon coupling strength.

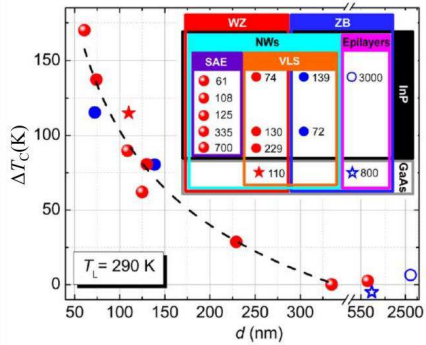

Fig. 4. Lattice and carrier temperature mismatch, $\Delta T_C$, experimentally extracted for 11 different InP and GaAs nanowires with various diameter, $d$, crystal morphology (WZ/ZB), and growth method (vapor-liquid-solid (VLS) vs metal-free selective-area epitaxy (SAE)), under continuous illumination with a 532 nm wavelength source. $T_L$ remains fixed at 290 K, while the power is varied in the range of 1-10⁴ W/cm². The dashed line is a hyperbolic fit to diameters<350nm. *Reproduced with permission from Nano Lett. 16, 3085 (2016). Copyright 2016, American Chemical Society.*

## IV.C. Experiments probing acoustic phonon dynamics

One way of probing the acoustic phonon dynamics is through its effect on macroscopic properties such as thermal conductivity, $k$. The thermal conductivity of a structure will depend on phonon parameters such as the density of states, the group velocity (both features of the phonon dispersion) and the mean free path of phonons that transport heat effectively.[113] There are numerous reports in the literature establishing that $k$ of nanowires decreases with smaller diameter. This has been observed in various nanowire materials platforms produced by various growth methods, for example Si,[114–116] InAs,[117–119] ZnO,[120] Ge/GeSi core-shell,[121] Si/SiGe superlattice,[122] and InSb[123]. Here we will briefly highlight some points regarding the $k$ in nanowires. Theoretical predictions based on phonon dispersion models for nanowires, which show modified branch structures and lowered group velocities, match the observations fairly well.[98,124–126] The decrease in $k$ is typically observed together with an increase in acoustic phonon boundary scattering (and less dominant Umklapp scattering).[114,115,120–122] An increased rate of boundary scattering is indeed to be expected as the diameter is decreased relative to the mean free path of acoustic phonons. Increased scattering of acoustic phonons increases their localization, and hence they transport heat less efficiently, resulting in greater temperature gradients, consistent with observa-tions, e.g. in Si nanoribbons.[127] In addition to a decrease of $k$ with smaller nanowire diameter, $k$ has been observed to decrease in nanowires with increasing density of hetero-interfaces,[122] and surface roughness[115,120], which, again, may be linked to increased phonon boundary scattering.

Since many of these reports include observations of a linear dependence of $k$ on diameter in nanowires with diameters as large as 100 nm or more, phonon confinement alone is not expected to explain this dependence (as discussed in Section III.B, phonon confinement effects are expected to play a role only for diameters smaller than about 10 nm). For example, a bulk phonon dispersion model was used to fit thermal conductivity measurements of Ge nanowires with a diameter of 62 nm, but it was found necessary to take phonon confinement effects into account to get a good fit for similar data taken from a Ge nanowire with 15 nm diameter.[121] It may also be added that theoretical work on Si nanowires between 1-8 nm in diameter, where confinement effects would be expected to be significant, did not see any drastic changes in thermal conductivity due to phonon confinement.[128,129]

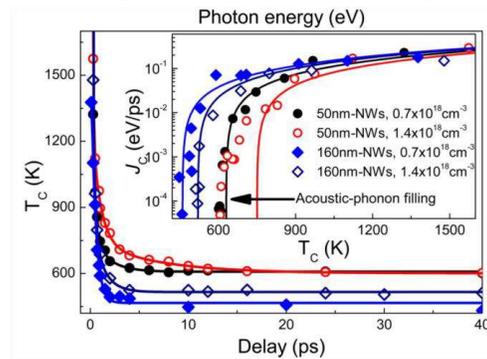

Fig. 5. Carrier temperature, $T_C$, as a function of time after excitation as extracted by transient photoluminescence measurements on InP crystal structure (WZ/ZB) superlattice nanowires. The inset shows the energy loss rate, $J_C$, as a function of $T_C$. Worth noting is that $J_C$ drops very steeply (many orders) at certain $T_C$, due to acoustic-phonon filling according to theory. The onset temperature is found to increase both with smaller nanowire diameter and with higher excitation density. *Reproduced with permission from Nano Lett. 13, 4280 (2013). Copyright 2013, American Chemical Society.*

Yong *et al.*[17] argue that they observe an acoustic phonon bottleneck (see Section III.A.3) by photoluminescence measurements at 300 K in InP nanowires containing a high amount of wurtzite (WZ)/zincblende (ZB) type-II interfaces. Separation of electron-hole pairs across the type-II interface is expected to lead to faster carrier relaxation and hence subsequent acoustic phonon generation, as well as additional acoustic phonon emission as electrons diffusing from WZ to ZB and relax. The authors predict that this leads to a fast build-up of acoustic phonons, resulting in the establishment of a non-equilibrium acoustic phonon distribution. Using transient photoluminescence with varying time delay, $T_C$ as a function of time after excitation pump was extracted for two different diameters and initial charge carrier densities (Fig. 5.). During the first couple of picoseconds, i.e. the timescale of carrier-LO phonon interactions, $T_C$ drops very rapidly, as the hot carriers equilibrate with the LO phonons. After this initial step, the cooling rate is seen to slow notably, to a point where $T_C$ is close to constant at the time scale of 40 ps (inset in Fig. 5.

shows the energy loss rate of hot carriers being drastically reduced). As this occurs at temperatures well above $T_L$ for most of the samples, it has been interpreted as an acoustic phonon bottleneck-effect, slowing down the carrier relaxation.[17] Notably, the $T_C$ at which the rate slows down is seen to increase with reduced diameter. It is approximately 150 K higher for the 50 nm diameter nanowire as compared to the 160 nm case. These results are in good agreement with those reported by Tedeschi *et al.*[16] in Fig. 4.. Yong *et al.*[17] note that the density of stacking faults (ZB segments) increased with decreasing diameter for the nanowires used in the study, and point to this as the possible source of the increasing $T_C$ for smaller diameters. In contrast, the inverse dependence of $T_C$ on nanowire diameter found by Tedeschi *et al.*[16] was observed independently of the nanowire growth method and crystal quality, pointing instead to the diameter as having the vital role. It would be valuable to devise studies where nanowire diameter and stacking fault density can be varied independently, in order to determine the respective roles of these variables in inhibiting the decay of hot carriers and increasing $T_C$.

### IV.D. Carrier-phonon interactions

Can energy-loss rates be reduced by manipulating electron-phonon interactions in nanowires? The interaction has two major factors that can be potentially engineered: (i) the phonon and electron dispersions (as seen for the thermal conductivity results, Section IV.C.), and (ii) the strength of the electron-phonon coupling. Taken together these two factors will determine the overall electron-phonon scattering rates and hence the energy loss rates. Reducing the strength of the electron-phonon coupling will mean a hot electron will be less likely to interact with, and emit, optical phonons. Existing theoretical and experimental work does not offer a clear picture, and reports exist of the electron-phonon coupling strength increasing, decreasing, and being independent of nanostructure size.[130] For a more in-depth discussion of this topic see e.g. the review by Kelley[130].

A small number of reports exists on measurements of the electron-phonon coupling strength in nanowires. For instance, an observed electronic heating effect across an InP heterostructure in InAs nanowires at low temperatures (1.2 - 4.2 K) was interpreted in terms of heat being transported interchangeably between electrons and phonons along the nanowire.[131] This interpretation required the assumption of a coupling strength between electrons and acoustic phonons three orders of magnitude stronger than previously predicted by theory.[131] Quantum wires formed at an InGaAs/InAlAs interface showed the energy loss rate per electron in the wire more than doubling compared to bulk, at electron temperatures of a few K. This could be indicative of an increased electron-phonon coupling strength, here ascribed to the singularities in the quasi-1D density of states.[132]

Conversely, more direct probing of the electron-phonon coupling strength by resonant Raman spectroscopy shows the electron-LO-phonon coupling strength decreasing with diameter in ZnO[133] and ZnTe[134] nanowires (Fig. 6). In both cases the decrease in coupling strength was associated with a decrease in the Fröhlich interaction. This is also consistent with several observations using resonant Raman spectroscopy on CdSe nanocrystals.[130] In that case, a smaller size is predicted to confine electrons and holes so that their wavefunctions spatially overlap, reducing the charge available for Coulombic interactions with the ions of the lattice (i.e. the Fröhlich interaction is reduced). A reduction of electron-phonon coupling has also been proposed to explain observations of a steady-state carrier temperatures well above the lattice temperature in InP and GaAs nanowires[16], and Si nanoribbons [127].

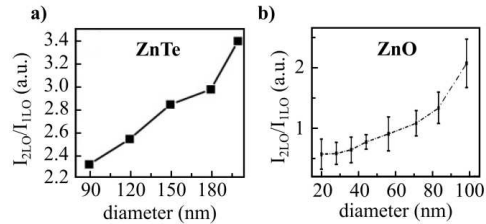

Fig. 6. The electron-LO phonon coupling strength can be quantified by comparing the intensity of the 2LO and 1LO Raman scattering peak, where a higher ratio $I_{2LO}/I_{1LO}$ relates to a stronger coupling, as detailed in ref.[134] Figures show $I_{2LO}/I_{1LO}$ as a function of nanowire diameter for (a) ZnTe nanowires, and (b) ZnO nanowires. *(a) Reproduced (adapted) with permission from Phys. Rev. B 85, 085418 (2012). Copyright 2012 by the American Physical Society (b) Reproduced (adapted) with permission from Phys. Rev. B 69, 113303 (2004). Copyright 2004, American Physical Society.*

## V. REALIZATION OF NANOWIRE HOT-CARRIER DEVICES

High-quality semiconducting nanowires are typically grown bottom-up using various epitaxial techniques, allowing for high control of material design as well as material savings.[20] The small diameter of the nanowire substantially relaxes the strain between layers with different lattice constant or crystal structure, allowing for much more flexible heterostructure engineering compared to bulk,[22] including atomically sharp interfaces.[21] In this section we will review the current state of experimental realizations of nanowire-based devices that make use of hot-carrier effects as an integral part of their design, with a special focus on HCPV. We will first, briefly, consider other related devices.

*Transistors.* Hot carriers are well known to have degrading effects in transistors.[9] As transistors get smaller, the associated electric fields get stronger, resulting in substantial generation of hot carriers. These hot carriers may get injected into the gate oxide and interface, altering the threshold voltage and gradually degrading the device. Efforts to mitigate the detrimental effect of hot carriers in nanowire based transistors include self-curing effects, realized by using either external annealing or built-in Joule heaters.[135] However, hot carriers may also offer advantages in transistors, such as high-speed operation based on ballistic transport. Efforts are underway to implement ballistic transistors in InAs nanowires, leveraging the advantages of flexible and dopant-free heterostructure design.[136] Using a tailored InAsP/InAs heterostructure[137] to launch ballistic, hot electrons of about 0.5 eV kinetic energy, in a three-terminal geometry, an estimated mean free path of 200-260 nm for the hot electrons was measured.[136]

*Plasmonic devices.* The typical photodetector is unable to detect light with energy less than the bandgap of the detector. By instead exciting hot carriers in a metal (plasmonic) structure, subsequently injecting them across a Schottky barrier into a semiconductor, carriers with energies below the bandgap can be detected. The principle of such a detector was demonstrated as early as 1888 by Stoletow,[138] and is today an active field of research for energy generation and sensing.[10–14] One advantage of this approach is that the energy spectrum of the hot carriers can to some degree be tuned by the design of the plasmonic resonance. Furthermore, energy extraction across a Schottky barrier at a metal-semiconductor interface allows the bandgap of the semiconductor to be chosen freely. Generally, such studies are performed using plasmonic metal elements on thin semiconductor films. Motivated by the large surface area on which the plasmonic particles can be deposited, nanowires have also been studied in this context.Hot-carrier injection from plasmonic gold particles on a nanowire surface, via Schottky contacts to nanowires of ZnO,[139] TiO$_2$,[140] and Cu$_2$O[24] has been observed. Regarding the efficiency of the photoemission process, White and Catchpole[141] note that since excitation in the metal can happen to a large extent well below the Fermi level, only a portion of excitations will result in carriers to be photo-emitted over the barrier. Strategies for increasing the extraction efficiency, by overcoming impedance mismatch, include roughening of interfaces and nanoscale shaping of plasmonic elements.[5,142,143]

For the remainder of section V we focus our discussions on the use of nanowire-based HCPV devices.

## V.A. The ideal hot-carrier photovoltaic device: motivation for using nanowires

Nanowires have several properties that make them particularly suited for the realization of HCPVs with high efficiency. To illustrate this, we begin by considering general requirements for an ideal HCPV device, as illustrated in Fig. 7. The device is characterized by an absorber region where a non-equilibrium concentration of photo-excited carriers is generated. We described these carriers by an elevated carrier temperature $T_C > T_L$, as well as a splitting of quasi-Fermi level $\mu_{n1}$ and $\mu_{p1}$ for electrons and holes respectively. Adjacent to the absorber region are two energy filters with energy-dependent transmission functions, $\Gamma_e(E)$ and $\Gamma_h(E)$, that allow the selective extraction of hot electrons and holes, respectively, into colder regions. The colder regions are here assumed to be at temperature $T_L$. The overarching aim of the operation of such a HCPV device is to achieve high, steady state, quasi-Fermi level splitting $\Delta\mu = \mu_{n1} - \mu_{p1}$ and temperature differential $\Delta T_C = (T_C - T_L)$. These two thermodynamic resources should be converted into electric current and voltage, with the highest possible electric power and efficiency.[144] Based on this general concept, one can identify the following general requirements for HCPV design:

*High carrier temperature differential.* Thermodynamically, the achievable energy conversion based on thermal extraction is limited by the Carnot efficiency $\eta_{CA} = \Delta T_C / T_C$. Thus, a high steady-state $\Delta T_C$ is required. Thermalization at a high $T_C$ requires a high density of hot carriers that can thermalize with one another (carrier-carrier scattering). This leads to the requirement of spatially confining the hot-carrier population within a region where ideally all excitation takes place (the absorber region in Fig. 7(b)).

*Low thermal conductivity.* Maintaining a high $\Delta T_C$ also implies minimizing lattice heat transported away from the absorber region, because a high acoustic phonon density in the absorber region is likely to contribute to slower carrier-lattice relaxation (as discussed in Section IV.C). $\Delta T_C$ has also been observed to increase exponentially with $T_L$ in nanowires (see Section IV.B).[16] Thus, maintaining a high $\Delta T_C$ might benefit from a low thermal conductivity in the host material.

*High precision band-engineering and optimized energy filters.* Careful heterostructure engineering is required across the entire device. Ideally, the absorber region has a narrow bandgap to absorb as many photons as possible, whereas the bandgap in the contact regions needs to be chosen to support a high operation voltage as well as current. Energy filters with highly selective transmission (both as a function of carrier type and energy) should allow for hot carriers to be emitted from the confined region, while maintaining a high $\Delta T_C$ across the filter. The energy filters should separate hot charge carriers at energies greater than this absorber bandgap, in order to achieve high operation voltage. Furthermore, the band design should ensure minimization of dark current through the energy filter (see Section V.B for a detailed discussion of energy filter design).

*Fast extraction.* The separation of hot carriers should take place before they have time to relax by transferring energy to the lattice. This requires energy filters to be positioned near the absorber region, on a similar length scale as the range of electron-phonon interaction (order of about 100 nm, see Table 1).

*Directional kinetic energy.* For high efficiency, the energy filter should be designed such that ideally all of the momentum of a carrier is directed towards the energy filter. This is a challenge in layered 3D devices where a hot carrier has on average only a third of its momentum in the same direction as extraction, and thus the other two thirds correspond to an unproductive heat loss from the absorber (for more on filter design, see section V.B.1).

Based on the above requirements, one can identify a number of strong reasons to consider nanowire-based HCPV, at least for fundamental studies and proof-of-principle devices:

- Perhaps most importantly, nanowires offer much higher freedom in heterostructure engineering than layered bulk devices, because strain relaxation allows the dislocation-free combination of lattice-mismatched semiconductors.[22] This makes nanowires an excellent platform for satisfying the complex band-engineering requirements indicated above, including the potential for advanced energy-filter design with very high precision and extremely sharp interfaces where desired.[21]

- Achieving a localized, high concentration of hot carriers is of vital importance in a HCPV device. The photonic properties of nanowires allow the tailoring of resonant optical modes to optimize local absorption (see Section V.C).[145–148] The high aspect ratio of nanowires makes them very suitable for the implementation of plasmonic elements that can be used to control absorption[149] or inject hot carriers.[24,139,140] Additionally, ordered arrays of nanowires have been demonstrated to possess a very low refractive index and significantly higher absorption than bulk or thin-film counterparts, mainly due to light scattering and trapping in and between the nanowires (Section V.D).[150–152]

- The high surface/volume ratio restricts the phase space for scattering, potentially affecting hot-carrier relaxation and phonon decay dynamics. In addition, thermal conductivity due to acoustic phonons can be strongly suppressed in nanowires, which may be one underlying reason for the high $T_C$ observed in nanowires under steady-state conditions (Sections IV.B and IV.C).[16]

- Finally, quasi-1D transport in nanowires comes with the advantage that most of the electrons momentum is focused in the forward direction. In fully 1D transport (wire diameter comparable to the electron wave length), this is known to enable much more efficient energy conversion by thermionic energy-filtering compared to layered 3D devices.[92] For intermediate wire diameters and a wide distribution of carrier energies, the role of wire diameter needs further exploration.

In the following, we discuss in more detail how these potential advantages of nanowires can be utilized to realize nanowire-based HCPV devices, and we review the current state of experiments.

## V.B. Hot-carrier extraction in nanowires.

### V.B.1. General considerations on energy filter design

The filters illustrated in Fig. 7 have two distinct roles. First, they are used to separate electron-hole pairs (carrier selectivity). Second, by selectively transmitting charge carriers with certain kinetic energy, they provide a mechanism to achieve an open circuit voltage larger than that of the Shockley-Queisser limit (energy selectivity).[33] In essence, this is possible because the energy filters work as thermoelectric energy converters, using the temperature differential $\Delta T_C$ to generate a thermoelectric voltage in addition to that provided by the quasi-Fermi level splitting $\Delta \mu$.[144,153]

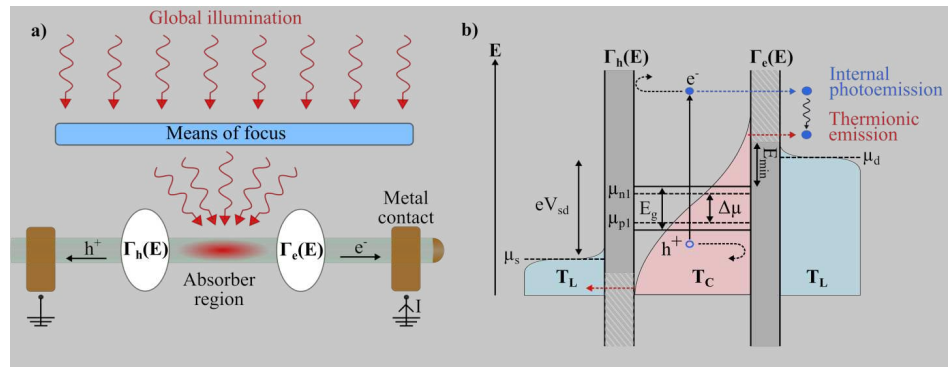

Fig. 7. (a) To realize an ideal HCPV device in a nanowire, light absorption should take place primarily within a specified absorber region (suggestions for how to realize this is discussed in Section V.C). The absorber region is enclosed by electron and hole filters described by transmission functions $\Gamma_e(E)$ and $\Gamma_h(E)$, respectively. Connecting the nanowire to metal source and drain contacts enables the device to generate a photocurrent. (b) Band diagram of absorber region and electron and hole filters (grey regions) designed to selectively transmit only one type of carrier. Photoexcitation of electron-hole pairs in the absorber (bandgap $E_g$) leads to a splitting $\Delta \mu = \mu_{n1} - \mu_{p1}$ of quasi-Fermi levels. Thermalization leads to a carrier distribution at $T_C > T_L$. Carriers can be extracted either before thermalizing (internal photoemission) or after (thermionic emission). Thermionic emission can occur above some threshold energy $E_{min}$. Extraction of hot carriers from the absorber regions, at temperature difference $\Delta T_C = (T_C - T_L)$, allows to add a thermoelectric voltage to $\Delta \mu$, ideally leading to an operation voltage $V_{sd} = (\mu_d - \mu_s)/e > \Delta \mu$ between drain and source contacts, where $e$ is the elementary charge.

Following photoexcitation, there are two different stages at which hot carriers can be extracted, as shown in Fig. 7(b). In one mechanism, referred to as *internal photoemission*, extraction takes place before hot carriers have undergone significant energy loss. This implies an extraction within about 10-100 fs of excitation, which, in a semiconductor, corresponds to ballistic transport over a distance of about 1-100 nm (Table 1). In a second mechanism, on a slower timescale, energy renormalization processes such as carrier-carrier scattering and impact ionization will lead to the formation of a thermalized distribution at temperature $T_C$. Extraction from the high-energy tail of this thermalized distribution is referred to as *thermionic emission*.

Which energy filter transmission function, $\Gamma(E)$, is optimal for HCPV devices? The answer to this question is complex and depends on material parameters, the light spectrum, and the design of the de-vice (including, in a self-consistent manner, the effect of the energy filters themselves on the achievable $T_C$, discussed more in Section VI). However, it is possible to offer some general considerations, which we briefly do in the following.

The overall aim is to extract as many of the photoexcited carriers as possible (to produce the highest possible current) against an operation voltage $V_{id}$ that is as high as possible. Extraction of completely non-thermalized electrons via internal photoemission across an energy barrier or Schottky contact with minimal transmission energy $E_{min}$ is generally not an efficient way of using their energy, because any kinetic energy above $E_{min}$ is lost via carrier relaxation. In particular, the excess energy could be put to better use by exciting carriers with low energy, allowing them to be extracted. For this reason, one can expect that the best strategy is to allow carriers to thermalize with one another before extraction.[154]

The objective to convert the kinetic energy stored in a quasi-thermalized, hot electron gas into electric current is very closely related to thermoelectric energy conversion. Specifically, there exists a body of literature[155–158] on how to optimize the electronic energy conversion efficiency, given two electron baths, one at a hot temperature (corresponding to $T_C$ in the present case), and one provided by a cold reservoir (corresponding to $T_L$). In the following we briefly review key insights from such studies.

When designing an energy filter for thermoelectric energy conversion, one faces a tradeoff between optimizing efficiency and power. Just like for any other type of heat engine, increasing one (e.g. efficiency) always comes at the cost of compromising the other (power in this case). It is, however, possible to find systems that have better tradeoffs than others. The maximum achievable electronic conversion efficiency from a thermodynamic perspective is the Carnot efficiency, in the case of a system such as Fig. 7(b) given by $\eta_{CA} = (T_C-T_L)/T_C$. Carnot efficiency can be realized by using an ideal energy filter with a delta-function shaped transmission function at a specific energy that depends on the temperatures and electrochemical potentials in the hot and cold reservoirs (Fig. 8(a)).[155,158] Importantly, the use of such an ideal energy-selective filter requires careful optimization and tuning of the position and the filter's selective energy window. In Section V.B.2 we describe how highly energy-efficient thermal-to-electric energy conversion for electrons has been achieved in nanowires.[27] Limpert et al.[144] showed that it is possible, in theory, to fulfill the conditions for Carnot efficiency simultaneously for two filters, one for electrons and one for holes. This makes it possible to use the thermoelectric voltage, resulting from $\Delta T_C$, to increase the open-circuit voltage above the Shockley-Queisser limit.[144] However, in spite of its thermodynamic efficiency, an ideal energy filter, with a delta-shaped transmission function,[155,158] has the severe drawback of infinitesimally small power production.[155] In principle, one can increase power by allowing for a transmission function with a finite energy window, for example one described by a Lorentzian with a width, $W_{FWHM}$ (Fig. 8(a)). But the trade-off is not very good, and efficiency in this case decreases drastically as power increases, because of back-flow of cold electrons through the low-energy tail of the Lorentzian. Nakpathomkun et al.[92] theoretically assessed the efficiency at maximum power production for a quantum dot energy-selective filter, for thermionic barriers embedded either into a 1D system (such as a thin nanowire), and for a 3D system. Fig. 8 shows the power and efficiency for various bias voltages at the chemical potential (kept constant) that delivers maximum power for each system. The efficiency at peak power was found to be $\eta_{max\,p} = 0.51\eta_{CA}$ for a quantum dot with a Lorentzian energy window with $W_{FWHM} << kT$ (Fig. 8(b)). Power production in a quantum dot is highest for $W_{FWHM} = 2.25\,kT$, but then with the low efficiency at peak power of $\eta_{max\,p} = 0.17\eta_{CA}$. For the 1D and 3D thermionic barrier (Fig. 8(d) and

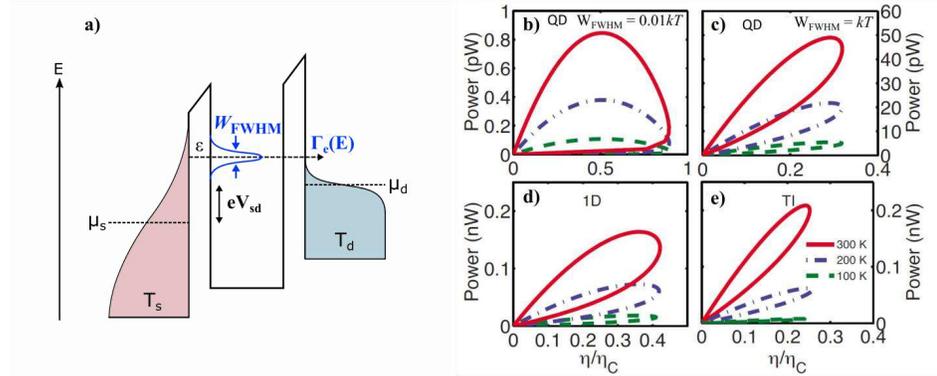

Fig. 8. (a) Principle of generation of a thermoelectric current using a resonant tunneling barrier (a quantum dot) as an energy filter. Temperature difference between source and drain, $\Delta T = T_s - T_d$ leads to a thermoelectric current of electrons via the resonant energy level, $\varepsilon$, with an energy width (full width at half maximum) $W_{FWHM}$, resulting in a voltage difference $V_{sd} = (\mu_d-\mu_s)/e$, where $e$ is the elementary charge. Trade-off between thermoelectric power and efficiency for different types of energy filter: (b) a quantum dot energy selective filter with energy window $W_{FWHM}$ = 0.01 $kT$, (c) a quantum dot filter with $W_{FWHM}$ = 1 $kT$ (d) a thermionic barrier embedded into a one-dimensional nanowire, and (e) a thermionic barrier embedded in a layered 3D structure. In each case, loop-like plots are obtained by scanning the system's bias voltage (corresponding to an external load) for constant electrochemical potential (chosen as the one that offers maximum power). In each case, a temperature differential $\Delta T$ = 30K was used at cold reservoirs temperatures of 300 K (red solid line), 200 K (blue dotted line), 100 K (green striped line). *Reproduced with permission from Phys. Rev. B 82, 235428 (2010). Copyright © 2010, American Physical Society.*

Fig. 8(e)), efficiencies of $\eta_{max\,p}$ = 0.36$\eta_{CA}$ and $\eta_{max\,p}$ = 0.24$\eta_{CA}$ can be achieved, respectively. Based on this, the 1D thermionic barrier offers the best power-efficiency trade-off.

More specifically for the case of HCPVs, Le Bris *et al.*[159] shows theoretically, taking into account hot-carrier relaxation (heat losses), that the drop in photovoltaic conversion efficiency (fraction of incident solar power converted) by increasing $W_{FWHM}$ saturates at levels still above 50% as $W_{FWHM} \gg kT$. The case $W_{FWHM} \gg kT$ would conceptually be the same as a thermionic barrier.

In conclusion, at the current state of HCPV device development, the best choice for an energy filter in quasi-1D nanowires appears to be a simple thermionic barrier. First, because of its favorable power-efficiency trade-off (Fig. 8(b)). Second, because it is far simpler to fabricate and operate than an energy-selective filter, which must be tunable to achieve its full potential.

### V.B.2. Energy-selective filter in a nanowire.

In 2018, Josefsson, Svilans *et al.*[27] demonstrated the conversion of heat stored in electrons in a nanowire into electricity with a performance very near to the ideal thermodynamic limits. To realize what they referred to as a quantum-dot heat engine, they used an InAs nanowire with two embedded InP segments serving as tunnel barriers that form a quantum dot (Fig. 9(a)). A temperature gradient was set up along the nanowire by electrically heating one end of it, establishing a cold and hot side of the quantum dot. A cryogenic system was used to cool the single-nanowire device to an operation temperature of about 1 K, in order to ensure a $kT$ much less than the spacing between quantum dot states.

Following the principles outlined in Fig. 8(a), energetic electrons were transported through the quantum dot at only one specific resonant level (quantum dot state). By tuning electrochemical potential using an external gate, and tuning bias voltage using an external load resistance, an electronic efficiency exceeding 0.7$\eta_{CA}$ was achieved under optimal conditions (Fig. 9(b)). Furthermore, when optimizing pow-er and the external load, an efficiency at maximum power of about $\eta_{max\,p} = 0.5\eta_{CA}$ was demonstrated. While the current extracted in this case was purely thermoelectric, and the power was low (order of fW) because of the very narrow energy filter used in order to maximize efficiency, the experiment demonstrates that highly efficient thermal-to-electric energy conversion can be achieved in nanowires in practice.

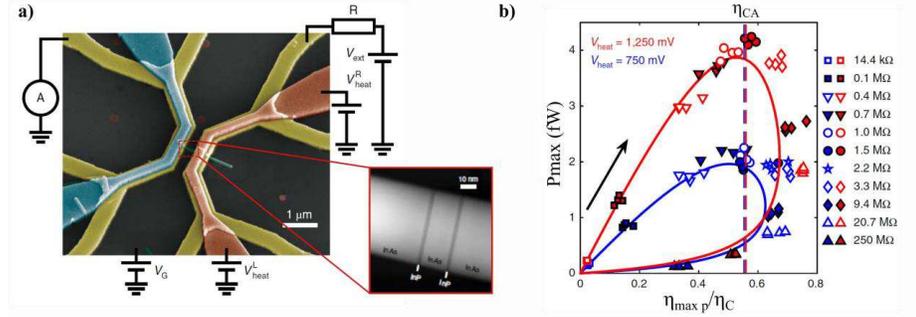

Fig. 9. (a) False-colored scanning electron microscope (SEM) image of a resonant tunneling barrier realized within a single InAs nanowire using two InP segments as a double-barrier to define a quantum dot (inset). The nanowire is contacted via Au metal contacts (yellow), allowing for current/voltage measurements and variation of an external load, $R$. A heating voltage, $V_{heat}$, is applied via top heaters[160] (red/blue for hot/cold), which are electrically insulated from the remaining device. The position of the resonant energy level can be tuned by applying a back-gate voltage $V_G$. (b) Maximum power, $P_{max}$, as a function of $\eta_{max\,p}$ for various $R$, under $V_{heat}$ of 1.250 mV (red) and 750 mV (blue). Marks indicate measurement and solid line theoretical prediction. *Reproduced with permission from Nature Nanotechnology 13, 920 (2018). Copyright 2018 Springer Nature.*

### V.B.3. Hot-carrier photovoltaic energy conversion in nanowires

Limpert *et al.*[74,145] demonstrated the use of heterostructure energy filters in single nanowires to separate photoexcited carriers. By globally illuminating contacted InAs nanowires containing an InP segment serving as a potential barrier (Fig. 10(a)), the generation of a current was observed. As supported by modeling (see Section V.C for more detail), this current generation was likely possible because of asymmetric light absorption with respect to the energy barrier, i.e. more absorption (and thus photogeneration of carriers) on one side of the barrier than the other (Fig. 10(d)). Nanowires containing a single InP barrier yielded open circuit voltages surpassing the Shockley-Queisser limit for the absorber material and light source used. Due to the asymmetry in effective mass of electrons/holes in InAs, as visualized in Fig. 1(a), a majority of the excess energy is expected to become kinetic energy of the electrons. Thus, the results of Ref.[74] can be seen as a strong indication that hot electrons are being extracted by the mechanisms illustrated in Fig. 10(a).[74] Studies of similar InAs nanowires by electron-beam induced current (EBIC) measurements confirm that the charge separation responsible for this current does indeed originate from the thermionic barrier (Fig. 10(b) and Fig. 10(c)).[161] The EBIC data demonstrates that carrier excitation close to the barrier is required for current generation, likely in order to prevent carrier relaxation before the electrons reach the barrier.

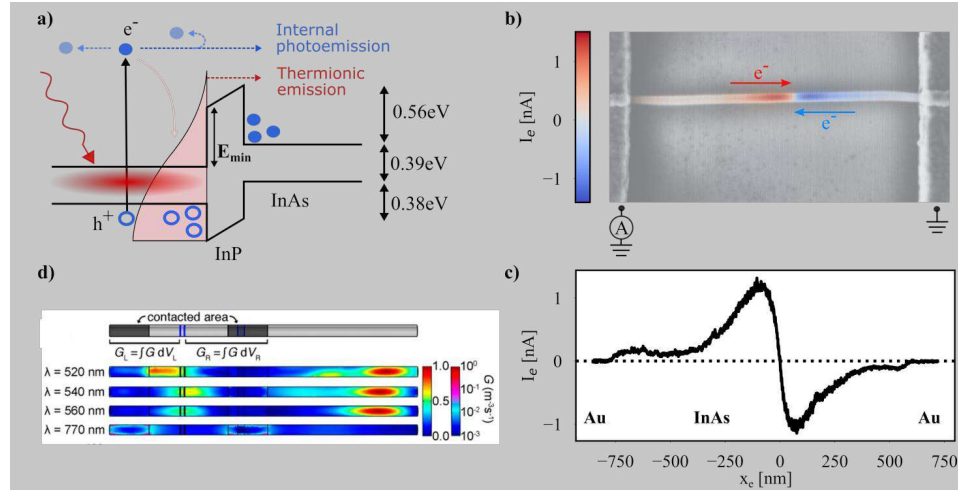

Fig. 10. (a) Band structure of an InP thermionic barrier embedded into an InAs nanowire. Hot electrons are excited on the left side of the barrier. Electrons transmitted across the barrier are separated from the respective holes, resulting in a net current. (b) Electron-beam induced current (EBIC) measurement on a contacted nanowire with composition as indicated in (a). The current changes polarity near the location of the InP segment, at $x_s = 0$ in the linecut of $I_e$ along the center of the nanowire (c). (d) Calculated (normalized) absorption rate, G, in InAs nanowire with metal (Au) contacts, under global, unpolarized, monochromatic illumination. The stripes between the contacted regions indicate the position of InP heterostructure energy filters. *(b) & (c) J. Fast, E. Barrigon, M. Kumar, Y. Chen, L. Samuelson, M. Borgström, A. Gustafson, S. Limpert, A. Burke, and H. Linke, Nanotechnology 31, 394004 (2020); licensed under a Creative Commons Attribution (CC BY) license. (d) Reproduced with permission from Nano Lett. 17, 4055 (2017). Copyright 2017 American Chemical Society.*

Using similar InAs/InP heterostructures (Fig. 11), Chen *et al.* controlled the wave-length of photoexcitation (using metal plasmonic elements, further discussed in Section V.B.1), and were thus able to selectively excite and extract hot carriers at energies either below or above the barrier height $E_{min}$, i.e. at energies above and below the points where internal photoemission is possible (Fig. 11 (d) and Fig. 11 (e)).[149] Two distinct behaviors of the short circuit current, $I_{sc}$, can be observed. For excitation of electrons to an energy below $E_{min}$ (Fig. 10(d)), current increases faster than linearly with illumination power, consistent with the expected thermionic extraction process. For excitation above barrier height (Fig. 10(e)), the total $I_{sc}$ increases linearly with power, consistent with internal photoemission. One can thus conclude that both, internal photoemission and thermionic emission, contribute to the current when a broad excitation spectrum is used. However, the thermionic contribution to $I_{sc}$ was one to two orders smaller in magnitude than that of photoemission, and thus the device appears to mainly operate on the basis of photoemission (Section V.C), which is not ideal for a HCPV device. In total, the internal quantum efficiency was estimated to 0.5-1.2% for the higher energies.[149] Based on the above studies one can extract estimates for the mean free path of hot electrons in InAs. From the EBIC results of Fig. 10(c), an effective hot-carrier relaxation length on the order of 100 nm was extracted.[161] Similar numbers have been found with the help of EBIC for holes in InSb/InAs nanowires.[162] Not yet published data of Kumar *et al.*[136] observed ballistic transport of 0.5 eV electrons over a distance on the order of 200-260nm in InAs nanowires. Chen *et al.*[149] observed lengths on the order of 30-40 nm under 1-1.3 eV illumination.[149] These values can be compared with the rough estimates of a characteristic length for carrier-phonon interaction on the order of 1 μm for electrons in InAs (Table 1), a notably higher value than observed. This may indicate that carrier-carrier scattering, with its shorter characteristic length, plays a strong role in the above experiments. We can conclude that significant extraction of hot electrons in a nanowire-system such as that of Fig. 10 appears to be possible, at best, within a few 100 nm of the excitation source.

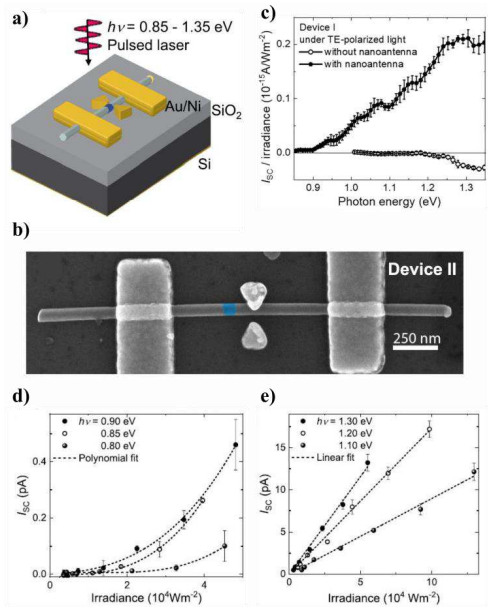

Fig. 11. Utilizing a "bow-tie" shaped plasmonic antenna made from Au to focus light absorption next to an energy filter embedded into an InAs nanowire. (a) Schematic of the device. (b) Scanning-electron micrograph of the finished device. Highlighted in blue is the InP segment that serves as energy filter. c) Experimentally measured photocurrent as function of illumination energy for a similar system. Performed under global illumination of TE polarized light (perpendicular to nanowire axis), before/after placing plasmonic antennas on the same device. (d) and (e) Short circuit current, $I_{sc}$, as a function of irradiance at a variety of excitation energies, the exponential/linear regimes are interpreted as thermionic emission/internal photoemission being the main source of the generated current. *Reprinted with permission from Nano Lett. 20, 4064 (2020). Copyright 2020 American Chemical Society.*

### V.B.4. Modelling of hot-carrier generation, transport and extraction in InAs-InP heterostructures.

Chen *et al.* estimated a quantum efficiency of their device (Fig. 11) around 0.5 – 1.2%, ewhen dominated by for internal photoemission.[149] To obtain a better understanding of what is limiting the quantum efficiency of the thermionic barrier, we have modelled the system of Fig. 10(a) using non-equilibrium Greens functions (NEGF). The simulations are based on a NEGF modelling framework for nanostructure-based photovoltaic devices, which includes the effects of interactions of charge carriers with photons and phonons in a picture of quantum transport ranging from the ballistic to the diffusive regime.[163,164] Simulations were carried out using a simple effective mass picture for the electronic struc-ture of electrons and holes. Charge transport was considered under monochromatic illumination and at zero applied bias voltage.

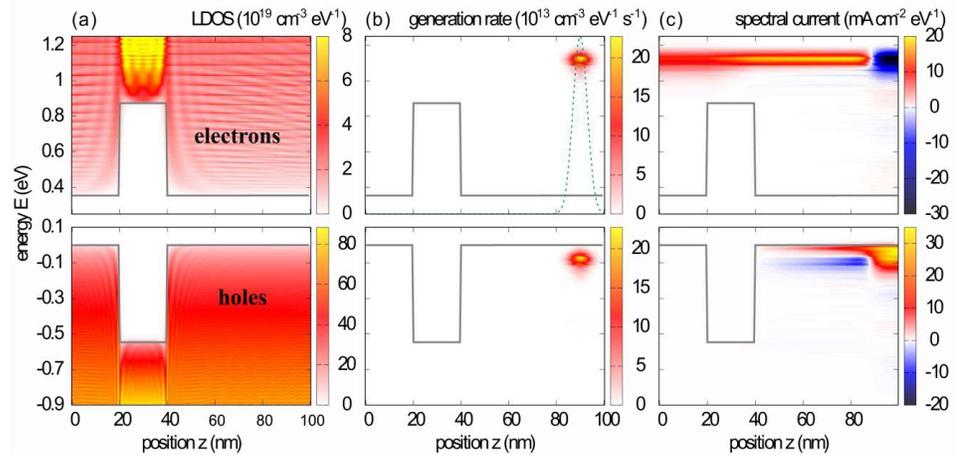

Fig. 12. NEGF simulation of hot-carrier photocurrent generation and extraction in InAs-InP heterostructures, electrons in the top windows and holes in the bottom windows. (a) Local density of states (LDOS). (b) Spectral generation rate for photon energy of 1.2 eV and localized Gaussian light intensity profile centered at 90 nm (dashed line). (c) Spectral current flow under illumination, including the effects of inelastic electron-phonon interaction.

Fig. 12(a) displays the local density of states of an InAs-InP heterostructure with a 20 nm long InP segment. There is a clear perturbation of the continuum above the InP barrier. The system is exposed to illumination of 1.2 eV in the form of a Gaussian light intensity profile centered at 90 nm (50 nm from the barrier), Fig. 12(b). Due to the disparity in effective mass of electrons and holes in InAs (electrons being lighter), the majority of the excess energy is absorbed as kinetic energy by electrons. The spectral current flow resulting from quasi-ballistic diffusion of this non-equilibrium carrier distribution is shown in Fig. 12(c), including the effects of inelastic electron-phonon interaction but not carrier-carrier interaction (thermalization). Positive (red) current is defined as electrons moving towards the left side (towards the barrier). While the hole flow is completely rectified by the InP barrier, the electron current exhibits both left and right travelling components as a portion of the electrons can make it over the barrier. Carrier relaxation due to scattering with phonons becomes notable on the left side of the barrier (where the spectral current becomes distributed over a larger energy range), but even there it does not appear to have a strong influence. It appears that the main limitation to transport over the barrier in this model is the transmissivity above the InP segment.

To better understand the system, we studied the energy dependence of the carrier generation and extraction process. More specifically, Fig. 13 shows (a) the generation rate, (b) the barrier transmission, and (c) the spectral current at zero in-plane quasi-momentum. It can be noted that at ideal, perfectly symmetric flat band bulk conditions, no more than 50% of the generated carriers will diffuse towards the barrier. The transmission in this case has been calculated for the case of ballistic transport, i.e. excluding any carrier relaxation processes. Notably, the transmission is seen to fluctuate strongly due to reflections on top of the barrier. These fluctuations clearly affect the resulting spectral current, e.g. with a dip at around 1.1 eV. This tells us that an additional factor of importance when considering a thermionic barrier HCPV device is how the barrier structure affects the transmission functions. For example, adding a smaller "pre-barrier" before a thermionic barrier may increase the transmission by around a factor of two.[165] Finally, we calculate the total photocurrent through the whole system as a function of excitation energy (with monochromatic light) in Fig. 13(d). The current is antisymmetric regarding the source position with respect to the barrier, and saturates at larger energies. This is in part due to the energy dependence of the absorption coefficient of InAs, but also due to the threshold for the internal quantum efficiency of 50% as half of the carriers diffuse away from the barrier. In spite of the over-simplified electronic structure used, this model reproduces well the behavior observed experimentally by Chen *et al.*[149], shown in the inset. However it can be noted that while the internal quantum efficiency observed there was on the order of ~1%, the NEGF models presented in this paper estimates an efficiency at least an order of magnitude larger. For a HCPV device ballistic extraction is not desired, however as internal photoemission appears to currently be the main source of extraction in the work by Chen *et al.* [149], this can also be seen as a sign that the efficiency could be notably increased by optimizing other losses in the system.

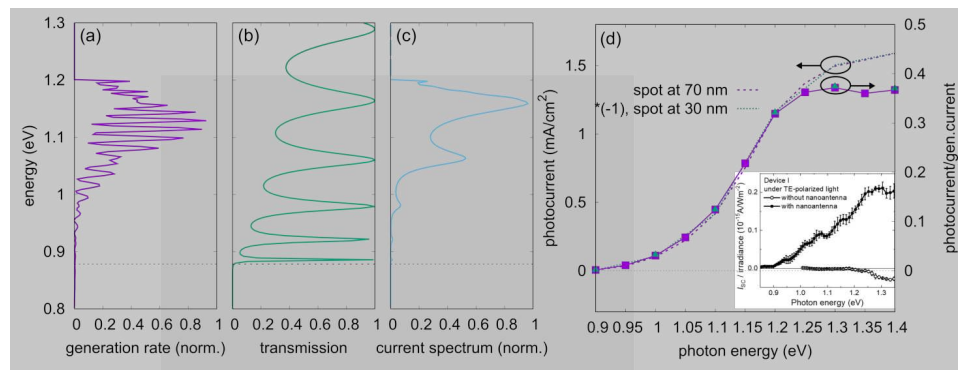

Fig. 13. Energy dependence of the carrier generation and extraction process: (a) Position-integrated spectral generation rate at zero transverse quasi-momentum (assuming no scattering). The dashed line denotes the top of the InP barrier. (b) Transmission function of the 20 nm InP barrier (for the case of ballistic transport). (c) Electron current spectrum at the left contact (z=0), reflecting the shape of both the generation rate and the barrier transmission function. (d) Dependence of the photocurrent on the energy of the monochromatic illumination. Inset shows experimentally measured[149] photocurrent of similar system, discussed in Section V.C and seen in full scale in Fig. 11 (c).

### V.C. Spatial control of light absorption in nanowires

One strong argument for using nanowires for HCPV experiments is that the one-dimensional geometry allows for several different means of tailoring absorption location. This is vital for a HCPV device for two reasons. First, optical absorption should only take place in a specified absorption region between hole and electron filter, ideally no more than a few 100 nm away (see Table 1 and Section V.B.2). Second, for efficient carrier thermalization a high carrier density (order of $10^{16}$ cm$^{-3}$) is typically required.[40,166] Achieving such conditions can prove quite challenging in three-dimensional absorbers. For example, no sizable electron-electron interaction has been observed in 100 nm thin film GaAs.[40] Because the typical nanowire dimensionalities are on the same scale as the wavelength of optical light, very strong resonant electromagnetic modes can form within them.[145–148] Optical absorption may therefore be distributed non-uniformly in a nanowire, with certain "hot spot" regions of high absorption at the location of wave mode maxima. The position of these maxima is determined by the geometry of the nanowire, its refractive index, and the excitation wavelength, commonly predicted by finite element modelling.[167]

Limpert et al. demonstrated, in the InAs/InP nanowire device discussed in section V.B.3 (Fig. 10), that absorption hot spots along a nanowire, in combination with an energy filter, can be used to generate a photocurrent.[145] Solutions to the Maxwell equations show well-identified wave mode maxima along the axial direction, with a location that depends on wavelength (Fig. 10(d)). Under global illumination, these devices generated a photocurrent, attributed to an imbalance of carriers being created on the two sides of the energy filter. Crucially, because the precise location of maxima shifted as function of wave length, also magnitude and directionality of the photocurrent varied with wavelength. The experimentally observed variation matched well with the predictions of absorption hot spots being located at different sides of the electron filter in the model. Whereas this experiment demonstrates the ability to use hot electrons to generate a photocurrent in an absorber material with very small bandgap (too small for a pn-junction solar cell to operate), it also had key limitations. Namely (i) the location of light absorption along the nanowire was not well controlled, and (ii) the presence of surface plasmon effects at the metal contacts in addition to photonic effects, clouding the picture (for example around 770 nm excitation in Fig. 10(d)).

Plasmonic antennas can be used to focus light absorption into sub-wavelength volumes. Specifically, coupling between the two metal antennas, such as in a bow-tie structure[168] with a tiny gap on the order of 10-100 nm, can lead to strongly enhanced optical field in the gap (for a review of plasmonic antennas, see e.g. Biagioni et al.[169]). Chen et al. placed Au bow-tie antennas around contacted single InAs nanowires, within 100 nm to one side of an InP segment, in order to focus light absorption very close to the thermionic energy filter (Fig. 11 (a) and Fig. 11 (b)).[149] The effect of the antennas to strongly enhance the observed short-circuit photocurrent $I_{sc}$ was convincingly demonstrated by measuring $I_{sc}$ on the same nanowire both before and after placing the plasmonic antenna (Fig. 11 (c)). These observations matches well with the mechanism for photocurrent generation described earlier (Fig. 10(a)) and the carrier separation observed[161] by EBIC in Fig. 10(b).

### V.D. Upscaling and optimization

We will end this section by a brief discussion on possibilities and challenges relating to how to go from proof-of-concept, single-nanowire experiments to real scale devices. Because HCPV devices are in part thermoelectric engines (see Section V.B.1), but the power of a single-nanowire heat engine is very small, it is useful to refer to considerations for the need to parallelize many nanoscale heat engines.[170] O'Dwyer et al.[171] expanded on Würfel's model[39] by modeling extraction with a finite number of 1D energy selective filters (such as in a nanowire array), and found that a very high density of filters (up to $10^{12}$ per cm$^2$) would be required in order to reach significant conversion efficiency. A similar conclusion was drawn by Kim *et al.*[172]

An obvious choice for achieving such high parallelization is the use of vertical nanowire arrays. Nanowire arrays come with the additional advantage that the array parameters, such as distance between nanowires (pitch), can be tuned to strongly enhance and control light absorption.[20] In fact, modeling shows that the optical generation of carriers in a nanowire array exposed to global illuminations is distributed asymmetrically through the wire, with the main portion of generation being focused within the top segment of the wire.[174] One could envision fabricating vertical nanowires with energy filters positioned around this region of major absorption, as visualized in Fig. 14(a). As discussed for single nanowires in the previous section, one could also envision the use of metallic plasmonic elements to achieve high absorption in a small region in the vicinity of energy filters (Fig. 14(b)).

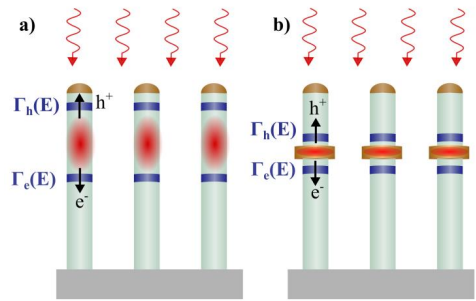

Fig. 14. Suggestions for two different approaches of realizing larger scale HCPV devices using nanowire arrays (a) Placing electron, $\Gamma_e(E)$, and hole, $\Gamma_h(E)$, energy filters around naturally occurring regions of higher absorption, controllable by array parameters such as pitch and wire diameter. (b) Placing plasmonic (metal) elements around each nanowire, focusing the absorption into narrow regimes of high excitation density, and placing energy filters on both sides of such region. Electric contacts can be added in the same way as for existing, pn-junction based nanowire array solar cells,[173] using transparent electrodes as a top contact, and the doped substrate as the bottom contact.

Just as for conventional pn junction photovoltaics, surface passivation will be of critical importance in nanowires, especially considering their high surface area.[20] Surface passivation is used to minimize the density and impact of surface states that provide additional recombination pathways, leading to high surface recombination velocities that hamper photovoltaic conversion efficiencies.[20] Passivation procedures are well developed for macroscopic devices, and much knowledge already exists for e.g. GaAs and InP nanowires, from efforts to realize efficient pn-junction nanowire solar cells.[20] For HCPV applications we are typically interested in narrow-bandgap materials, such as InAs ($E_g \approx 0.4$ eV). Surface oxidation of InAs nanowires has been shown to introduce a large number of trap states at energies above the band edge, where hot electrons may get trapped. Illumination of such samples will cause hot electrons to be trapped, resulting in a negative photoconductivity response.[175,176] Successful passivation of single InAs nanowires by a shell of Al$_2$O$_3$ has been reported.[175] It can also be noted that deposition of oxides such as Ta$_2$O$_5$,[177] ZnO,[178] and TiO$_2$,[179] on top of InP thin film solar cells have been demonstrated to provide pas-sivation (reducing surface recombination) and to serve as carrier-selective contacts due to the band off-sets. In nanowire devices where radial extraction is considered, this approach could provide a useful means of contacting and passivating at the same time, warranting further investigation.

## VI. OUTLOOK AND DISCUSSION

We have reviewed the literature relating to hot-carrier dynamics in nanowires with a focus on the use of nanowires for future hot-carrier photovoltaics (HCPV). A number of promising results have already been obtained using nanowires, such as: enhanced carrier temperatures compared to bulk (Section IV.B), hot-carrier separation at large open circuit voltage (Section V.B.3), and the designing of high-performing filters for hot-carrier extraction, enabling the harvesting of electronic heat with efficiency near thermo-dynamic limits (Section V.B.2). However, a number of questions still need to be answered before the potential of nanowires for high-efficiency HCPV can be fully evaluated. In the following, we briefly dis-cuss key, open questions.

*What is the mechanism for enhanced $T_C$ under steady-state illumination in nanowires, and can $T_C$ be further increased?* The experiments by Tedeschi *et al.*[16] (see Section IV.B) show a drastic, systematic in-crease of $T_C$ for decreasing nanowire diameter, impressively demonstrating that nanowires may have fundamental advantages over bulk materials for HCPV. However, the mechanism for this increase in car-rier temperature needs to be better understood. Currently, the most likely explanation is a phonon-bottleneck effect related to the reduced thermal conductivity of the nanowires, caused by increasing phonon boundary scattering with decreasing diameter (Section IV.B.C). To further explore the mecha-nism responsible for these observations, one suggestion is to study carrier relaxation and thermal con-ductivity for the same sample while tuning parameters expected to impact boundary scattering. For ex-ample, this could be done by repeating the photoluminescence studies reported by Yong *et al.*[17], but keeping the nanowire diameter constant while varying the density of WZ/ZB stacking faults. More work is needed before the full potential of nanowires for enhancing hot-carrier temperatures and lifetimes can be comprehensively evaluated.

*How can the quantum efficiency of HCPVs in nanowires be further improved?* Hot-carrier extraction in nanowires has been demonstrated, including the observation of a thermionic current and an open circuit voltage above the Shockley-Queisser limit[74] in InAs/InP nanowires.[180] However, the quantum effi-ciency in existing devices is still estimated to be small, on the order of 1% or less,[180] and further engineer-ing is required to reduce losses. Crucially, the NEGF results presented in this work (Section V.B.4) indicate that the current experimental limitations are not of a fundamental nature, and could be substantially increased. For example, the internal quantum efficiency may be increased by designing the barrier shape in order to minimize carrier reflection, such as adding pre-barriers.[165] Perhaps the most obvious next step in experiments would be to realize a conventional HCPV structure, as described in Fig. 7(b) of Sec-tion V.A, with carrier selective filters for each carrier type (i.e. one filter for electrons and another for holes). As a first step, it is likely sufficient to focus on the carrier type that receives most of the excess energy (typically electrons), and use carrier selective filters to block these from flowing away in other directions. It might be instructive in future theory and experiment to evaluate whether having two energy barriers, one for each carrier type, is even expected to be beneficial.

*Can we actually assume that hot carriers thermalize with each other?* As discussed in e.g. Section III.1 and V.A, it seems very likely that efficient HCPVs require the ultrafast establishment of of a thermal quasi-equilibrium among carriers, whereby energy is exchanged between high- and low-energy carriers. In view of the findings for bulk[40] and QW systems,[166] this would require charge carrier densities in excess of $10^{16}$ cm$^{-3}$. While nanowires offer the possibility of focusing light into narrow spots with correspondingly large local generation rate,[180] the carriers are essentially generated in extended electronic states. This extension limits the enhancement of carrier-carrier scattering rates in the way which is possible in narrow quantum well[2] and quantum dots[1]. As opposed to the situation in nanowires under pulsed laser excitation,[180] the attainability of a sufficiently thermalized carrier population is thus uncertain for applications operating under steady-state generation, even at the highest concentrations of sunlight. On a more practical note, high illumination intensity may consequently lead to strong heating of the lattice. In combination with the reduced thermal conductivity of nanowires, this heating could have detrimental impact on device performance and integrity. There is a need for additional work on the feasibility of forming an actual thermal distribution - or, at least, redistributing the energy efficiently - under solar spectrum illumination. More insights on the type of carrier distribution that can be expected under these illumination conditions will then inform efforts to optimize the hot-carrier extraction process, as discussed in subsequent sections. Note that while a thermal distribution in quasi-equilibrium as considered throughout most of this article can only be expected in the absence of an electric field the presence of an electric field does not hinder the extraction of hot carriers (but it does further complicate the description of the distribution).

*How can a HCPV device be optimized on a system level?* For a future, optimized HCPV device, careful optimization of the entire system is required. As a tool to help envisage how this might be done, we illustrate a simple particle- and energy balance in Fig. 15(a). As a system, we consider the photo-excited electrons in the absorber region of a HCPV, as illustrated in Fig. 7. In steady state operation, the flows of energy and particles into this system are balanced by respective flows of energy and particles out of the system. Similar rate models have been proposed in several studies, see e.g. ref.[56,181–184].

Beginning with the particle balance (Fig. 15(a)), the relevant particle flow rates are:

- $\Gamma_{gen}$, the rate at which carriers are generated by photoexcitation.
- $\Gamma_{rec}$, the rate of carrier recombination.
- $\Gamma_{ext}$, the rate at which carriers are extracted via the energy filters.

These particles flows will be accompanied by a quasi-Fermi level splitting, $\Delta\mu$. Considering the energy balance for the system (Fig. 15(b)), one needs to additionally consider:

- $\int \Gamma_{c\text{-}ph}(E)dE$, the energy flow rate at which the system loses heat to the lattice by carrier-phonon scattering.
- $\Gamma_{c\text{-}c}$, the rate of carrier-carrier scattering, contributing to the formation of $T_c$. Since this process only re-distributes energy among carriers, it is not associated with a flow across the system boundary.

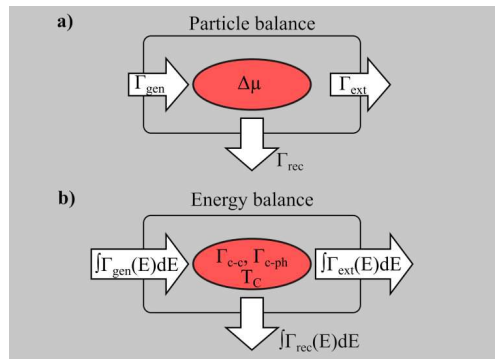

Fig. 15. Illustration of (a) particle balance and (b) energy balance of the hot electron gas in the absorber region of a HCPV device, such as illustrated in Fig. 7. The individual rates shown are defined in the main text.

Each of the above rates is energy dependent and integration over all energies is required for a full picture. The energy flows in and out of the system, in combination with the energy redistribution within the system, will determine the $T_C$ that can be maintained. Balance equations between these rates can then be used, for example, to find the operation point of a HCPV device.[185] Here, we aim to use this illustration as a help for discussing open questions in nanowire HCPV design.

The end goal of HCPV design is to maximize both output power and conversion efficiency. Both are assisted by a high open circuit voltage, which results from a combination of $\Delta\mu$ (similar to a conventional solar cell) and the additional thermoelectric voltage enabled by utilizing $\Delta T_C$.[144] The voltage contribution from $\Delta T_C$ is what sets HCPV apart from standard PV, and thus we here choose to focus on its optimization.

For optimal $\Delta T_C$, $\Gamma_{c\text{-}c}$ should be faster than any other rates so that a thermal population can be established, where high energy carriers redistribute part of their excess energy to low energy carriers.[15] A particle balance in steady state requires that $\Gamma_{gen} = \Gamma_{ext} + \Gamma_{rec}$ (Fig. 15(a)). If one can achieve a situation where $\Gamma_{ext} \gg \Gamma_{rec}$, then one can approximate $\Gamma_{gen} \approx \Gamma_{ext}$. To minimize energy losses (Fig. 15(b)), extraction must occur faster than the energy loss rates, $\Gamma_{ext} \gg \Gamma_{c\text{-}ph}$, $\Gamma_{rec}$. Based on these basic considerations, one can make the crude conclusion that the desired situation is: $\Gamma_{c\text{-}c} \gg \Gamma_{ext} \approx \Gamma_{gen} \gg \Gamma_{c\text{-}ph}$, $\Gamma_{rec}$.

However, these basic considerations neglect important and non-trivial interdependencies. For example, if $\int\Gamma_{ext}(E)dE$ is of the same order of magnitude as $\int\Gamma_{gen}(E)dE$, this will necessarily cool the carrier temperature, $T_C$, in turn modifying the other particle and energy flow rates. These interdependencies make the electron filter design a highly non-trivial system consideration (see also below). We identify the optimization of such a system as one of the main open questions that should be addressed in future work to assess the potential of any HCPV system.

*What is the ideal structure for energy filter?* The ideal design of an energy filter for HCPV applications is an open question and requires system optimization, as highlighted in the discussion in the preceding question and in Fig. 15. In short, the ideal HCPV energy filter needs to (i) allow for high quantum efficiency and high current at high voltage, (ii) prevent excessive cooling of the hot carrier gas (maintain a high $T_C$) in interplay with all other cooling rates, and (iii) be optimized (high efficiency at maximum power) for the actual operating point (illumination power, external load) of the device. Such an optimal filter will likely be neither a simple energy-selective filter (Fig. 8), which limits power flow, nor a simple thermionic barrier (Fig. 10), which allows all high-energy electrons above some threshold energy to escape. Optimization of a complete system becomes possible first when our understanding of the hot-carrier generation process allows us to design a complete system. At the current stage, and summarizing the discussions in Section V.B, we suggest that carrier extraction via a 1D thermionic barrier appears the best option for experiments in nanowires.[92] It offers the advantages of (i) a good trade-off between power and efficiency (Section.B.2), (ii) barrier shape and height are the only parameters to tune for optimization, and (iii) simple and scalable fabrication. However, the power-efficiency trade-off of a thermionic barrier embedded into nanowires of finite diameter (where transport is not completely one-dimensional) requires further investigation.

*What are our recommendations for a nanowire HCPV system?*
There are many factors to consider when designing an ideal HCPV system. The nontrivial interplay between hot carrier dynamics, control over optical absorption, and band-structure engineering for carrier and energy-selectivity makes it challenging to today point out a specific nanowire system that may be ideal for HCPV. Based on the literature discussed in this article we however offer some general considerations for what such a system might include.

- While the underlying mechanism is not yet clear, there exists strong evidence that $T_c$ increases with decreasing nanowire diameter as well as with increasing density of stacking faults.[16,17] This relation is also often accompanied by a decrease in $k$,[114–123] which, in addition, has been observed to decrease with the roughness of nanowire surface.[115,120] As such, it seems likely that an ideal nanowire HCPV device would take advantage of one or several of these effects.

- As for carrier- and energy-selective filters, it seems reasonable to at this stage focus on thermionic barriers for the simplicity of fabrication, good trade-off between power and efficiency for a thermoelectric system,[92] and relatively simple optimization. To increase the transmission of such one could consider integrating stepwise or graded barriers.[165] Due to the electron-hole asymmetry of the effective mass in most III-V materials we believe that it is currently sufficient to focus on energy-selectivity only for electrons, and to focus solely on carrier selectivity when it comes to holes.

- As for material choices, low band gap and low $k$ are expected as basic requirements for the absorber material. For now, it seems reasonable to continue focusing on the well-known III-V materials with demonstrated hot-carrier effects such as InAs, InP, GaAs, and their alloys. Previous theoretical investigations, that aimed to identify materials for bulk–like hot carrier absorbers, identified a large size mismatch between anion and cation atoms in polar materials as a selection criterion.[186,187] Subsequent experimental results for group III nitride compounds (III-Ns), such as InGaN, support this conclusion,[57] suggesting further investigations of III-N based nanowires might be warranted.

*Can nanowire-based HCPV devices be scaled up?* A possible pathway towards upscaling of nanowire-based HCPV devices has been presented in Section V.D, building on substantial work already done for pn-junction based nanowire array solar cells. More detailed investigations, both by modeling and by experiment, are required to better understand the potential specifically for HCPV. For example, more theoretical work can contribute to fully understanding the potential of photonic engineering to achieve absorption hot spots in nanowire arrays (Fig. 14), and the expected impact on device performance. Further studies of passivation on narrow-bandgap material nanowires may also be of great utility for improving carrier collection efficiency. For instance, it is well established that bare InAs and GaAs nanowires have much higher surface recombination rates than bare InP nanowires,[188] and passivated surfaces improve carrier lifetimes and light emission.[189]

In summary, nanowires display a number of intriguing properties directly relevant to hot carriers. These properties make nanowires a great platform for fundamental and proof-of-principle experiments in HCPVs. There are, however, a number of open questions that require further investigation to answer the question of whether they are a suitable platform for proof-of-performance devices. More experimental and theoretical work is required before a complete picture of the operation of a nanowire-based HCPV device can be outlined. In this article we have summarized the current progress and made suggestions of where future work should be directed, in order to allow for a full evaluation of the achievable performance of nanowire-based HCPVs.

**Acknowledgements**

The authors acknowledge financial support by the Knut and Alice Wallenberg Foundation (project 2016-0089), support from the Australian Research Council through Discovery Project DP170102677.

**Data availability**

The data that support the findings of this study are available within the article.